\documentclass[a4paper,fleqn,usenatbib]{mnras}
\usepackage[T1]{fontenc}
\usepackage{ae,aecompl}

\usepackage{amsmath}
\usepackage{epsfig}
\usepackage{graphicx}
\usepackage{lscape}
\usepackage{subfig}

\usepackage{amssymb}
\usepackage{color}

\newcommand {\bc}{\begin {center}}
\newcommand {\ec}{\end {center}}
\newcommand {\be}{\begin {equation}}
\newcommand {\ee}{\end {equation}}
\newcommand {\beq}{\begin {eqnarray}}
\newcommand {\eeq}{\end {eqnarray}}

\def\flux{erg s$^{-1}$ cm$^{-2}$}
\def\lum{erg s$^{-1}$}

\def\swift{Swift~J0243.6+6124}

\title[On the magnetic field in Swift~J0243.6+6124]
{On the magnetic field of the first Galactic ultraluminous X-ray pulsar Swift~J0243.6+6124}

\author[S.~S.~Tsygankov et al.]
{Sergey~S.~Tsygankov,$^{1,2}$\thanks{E-mail: sergey.tsygankov@utu.fi; juri.poutanen@utu.fi}
Victor Doroshenko,$^{3}$
Alexander A.~Mushtukov,$^{4,2}$\newauthor 
Alexander A.~Lutovinov$^{2}$
and Juri~Poutanen$^{1,2,5}$
 \\
$^1$Tuorla Observatory, Department of Physics and Astronomy,  FI-20014 University of Turku, Finland \\
$^2$Space Research Institute of the Russian Academy of Sciences, Profsoyuznaya str. 84/32, Moscow 117997, Russia \\
$^3$Institut f\"ur Astronomie und Astrophysik, Universit\"at T\"ubingen, Sand 1, D-72076 T\"ubingen, Germany \\
$^4$Anton Pannekoek Institute of Astronomy, University of Amsterdam, Science Park 904, 1098 XH Amsterdam, The Netherlands\\
$^5$Nordita, KTH Royal Institute of Technology and Stockholm University, Roslagstullsbacken 23, SE-10691 Stockholm, Sweden}

\date{Accepted 2018 June 22. Received 2018 June 18; in original form 2018 April 22}

\pubyear{2018}

\begin{document}
\label{firstpage}
\pagerange{\pageref{firstpage}--\pageref{lastpage}}
\maketitle

\begin{abstract}
We report on the monitoring of the final stage of the outburst from
the first Galactic ultraluminous X-ray pulsar \swift, which reached
$\sim$40 Eddington luminosities.  The main aim of the monitoring
program with the {\it Swift}/XRT telescope was to measure the magnetic
field of the neutron star using the luminosity of transition to the
``propeller'' state.  The visibility constraints, unfortunately, did
not permit us to observe the source down to the fluxes low enough to
detect such a transition.  The tight upper limit on the propeller
luminosity $L_{\rm prop}<6.8\times10^{35}$~\lum\ implies the dipole
component of the magnetic field $B<10^{13}$~G. On the other hand, the
observed evolution of the pulse profile and of the pulsed fraction
with flux points to a change of the emission region geometry at the
critical luminosity $L_{\rm crit}\sim3\times10^{38}$~\lum\ both in the
rising and declining parts of the outburst.  We associate the observed
change with the onset of the accretion column, which allows us to get
an independent estimate of the magnetic field strength close to the
neutron stars surface of $B>10^{13}$~G.  Given the existing
uncertainty in the effective magnetosphere size, we conclude that both
estimates are marginally compatible with each other.

\end{abstract}


\section{Introduction}
\label{intro}

The transient X-ray pulsar \swift\ was discovered by the {\it
  Swift}/XRT telescope \citep{2017ATel10809....1K} after an alert
generated by the MAXI monitor on 2017 September 29
\citep{2017ATel10803....1S}. Spin period of the neutron star (NS) was
measured at 9.68\,s \citep{2017ATel10809....1K,
  2017ATel10812....1J}. 
 Analysis of the observed spin frequency
changes allowed also to determine orbital parameters of the system
using the {\it Fermi}/GBM data, which revealed a low-eccentric orbit
($e\sim0.1$) with period $\sim$28 d
\citep{2018A&A...613A..19D,2018ATel11280....1J}.

Based on the optical spectroscopic observations,
\cite{2017ATel10822....1K} suggested that optical counterpart in the
system is late Oe- or early Be-type star.  The Be/X-ray binary (BeXRB)
nature of the source was later confirmed by
\cite{2017ATel10968....1B}. With peak flux reaching $F_{\rm
  peak}\sim7\times10^{-7}$ \flux\ in the outburst maximum, the source
represents one of the brightest X-ray sources and the brightest BeXRB
ever observed.  The distance to the source was independently estimated
using the X-ray and optical observations. Optical photometry and
spectroscopy obtained from the RTT-150 and BTA telescopes allowed to
estimate the mass $M_{\rm opt}=16\pm2$~M$_{\odot}$ and radius $R_{\rm
  opt}=7\pm2$~R$_{\odot}$ of the Be counterpart, which implies the
distance to the source of about 2.5\,kpc
\citep{2017ATel10968....1B}. On the other hand, analysis of the
pulsars spin-up rate as a function of luminosity performed by
\cite{2018A&A...613A..19D} implies a lower limit on distance of
$\sim5$\,kpc.  This disagreement was finally resolved by the {\it
  Gaia} observatory measured distance to the system of
$d=7.3^{+1.5}_{-1.2}$~kpc\footnote{obtained as a median value,
  assuming the priors recommended by the {\it Gaia} team, with the
  errors given at 16th and 84th percentile} (van den Eijnden et al.,
in prep.).  Such a distance implies peak luminosity of up to
$\sim5\times10^{39}$ \lum, exceeding the Eddington limit for a NS by a
factor of 40.  Thus, \swift\ can be considered as the first Galactic
X-ray pulsar, belonging to the recently discovered family of
ultraluminous X-ray pulsars \citep[see
  e.g.,][]{2014Natur.514..202B,2016ApJ...831L..14F,2017MNRAS.466L..48I,2017Sci...355..817I,2017A&A...605A..39T}.
It is worth mentioning that here and below we refer to isotropic
luminosity. Although some beaming of emission can be expected in the
case of XRPs, we argue that for \swift\ it is negligible. It follows
from the smoothness of the source lightcurve, which does not show any
features even when the pulse profile changes significantly, i.e. when
the critical and the Eddington luminosities are passed and the beaming
pattern is changed (see below).

\swift\ was observed with the {\it NuSTAR} observatory several times
throughout the outburst. Preliminary results of the X-ray broadband
spectroscopy demonstrate the presence of a high-temperature black body
component ($kT\sim3$ keV) in addition to the typical for X-ray pulsars
cut-off power-law with photon index of $\sim$1 and $E_{\rm cut}\sim20$
keV \citep{2017ATel10866....1B,2018MNRAS.474.4432J}. The X-ray
spectrum of the source revealed no evidence for additional absorption
features which could be associated with the cyclotron resonant
scattering and provide an estimate of the magnetic field of the NS
\citep[for a review see e.g.,][]{2015A&ARv..23....2W}.

Knowledge of the magnetic field is essential for understanding
physical processes responsible for the observed behaviour of X-ray
pulsars, such as their extreme luminosities, significantly exceeding
Eddington limit \citep[][]{2015MNRAS.454.2539M}. The magnetic field
strength in \swift\ was estimated based on the observed spin-up rate
of the pulsar and points to above-average value of $B\sim10^{13}$\,G
\citep{2018A&A...613A..19D} under assumption that the source is
located at distance $\sim$6.6\,kpc. While this estimate is in line
with non-detection of the cyclotron absorption line in the wide energy
range 0.3--80\,keV, it is not model independent and thus requires
verification.

Here we attempt to independently estimate the magnetic field of the NS
based on observation of the transition of the source to the so called
``propeller regime'' when the accretion is centrifugally inhibited by
the rotating magnetic field lines \citep{1975A&A....39..185I}. This
method was already verified in a very broad range of magnetic field
strengths using a sample of magnetized NSs, consisting of accreting
millisecond pulsar, classical X-ray pulsars and even pulsating
ultra-luminous X-ray source
\citep{2016MNRAS.457.1101T,2016A&A...593A..16T,2017ApJ...834..209L}. In
the case of \swift\ its relative proximity and very good coverage of
the tail of the outburst with the {\it Swift}/XRT telescope allowed us
to monitor the source down to very low luminosities and to put an
upper limit on the dipole component of the NS magnetic field in the
system. Unfortunately, continuation of the monitoring until an actual
transition to the propeller regime was impossible due to the Sun
constraints of the visibility.

\section{Data analysis}
We mostly rely on the data obtained with the XRT telescope
\citep{2005SSRv..120..165B} onboard the {\it Neil Gehrels Swift
  Observatory} \citep{2004ApJ...611.1005G} during the fading phase of
the outburst. Depending on the source brightness the observations were
performed both in Windowed Timing (WT) and Photon Counting (PC)
modes. The data reduction, i.e. spectrum extraction, was done using
the online tools
\citep[][]{2009MNRAS.397.1177E}\footnote{\url{http://www.swift.ac.uk/user_objects/}}
provided by the UK Swift Science Data Centre.  Only zero grade (single
pixel) events were included into the product.

\begin{figure}
\centering
\includegraphics[width=0.98\columnwidth, bb=55 265 575 680]{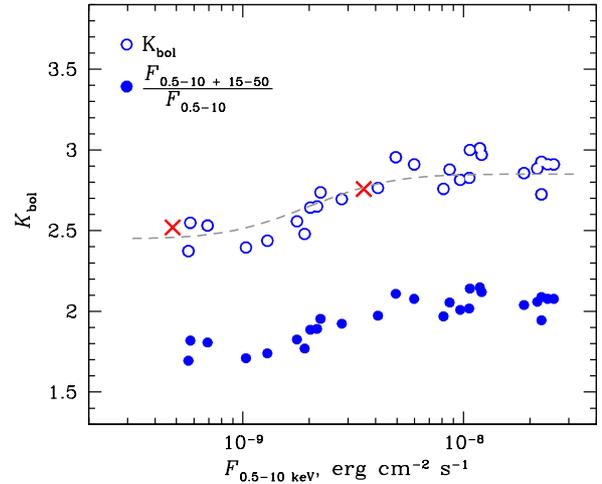}
\caption{Dependence of the bolometric correction factor $K_{\rm
    bol}$ on the observed source flux in the 0.5--10 keV energy band (open
  circles), obtained after MJD 58110. Filled circles show the ratio of
  the total flux in the 0.5--10 keV plus 15--50 keV energy bands to the 0.5--10 keV
  flux as a function of the 0.5--10 keV flux. Red crosses show the
  $K_{\rm bol}$ values obtained from the broad-band spectra collected
  with the {\it NuSTAR} observatory. Grey dashed line shows the best fit to
  the $K_{\rm bol}$ values with the Gauss error function.}\label{fig:bol}
\end{figure}

The source flux in each observation was determined based on the
results of spectral fitting in the {\sc XSPEC} package assuming the
absorbed power-law model.  Taking into the account the low counting
statistics in the very end of the outburst, we binned the spectra to
have at least 1 count per energy bin and fitted them using W-statistic
\citep{1979ApJ...230..274W}.\footnote{see {\sc xspec} manual;
  \url{https://heasarc.gsfc.nasa.gov/xanadu/}
  \url{xspec/manual/XSappendixStatistics.html}} Given the known
calibration uncertainties at low
energies,\footnote{\url{http://www.swift.ac.uk/analysis/xrt/digest_cal.php}}
we restricted our spectral analysis of the WT data to the 0.8--10 keV
band and 0.3--10 keV in the PC data. In our spectral analysis we
avoided usage of the {\it Swift}/XRT data for fluxes above
$\sim10^{-8}$~\flux\ strongly affected by the pile-up. Therefore our
dataset starts from MJD~58110. This restriction does not affect any of
our conclusions as only the final phase of the outburst is relevant
for detection of the transition to the propeller regime.

To estimate the source luminosity and the mass accretion rate, and
thus the magnetospheric radius of \swift\ (see below), a reliable
estimate of the bolometric flux is required. We estimate it based on
the observed {\it Swift}/XRT fluxes following the procedure described
in \cite{2017A&A...605A..39T}. The observed spectra of X-ray pulsars
are known to change with luminosity, so the bolometric correction is
also luminosity dependent.  To account for that we used the ratio of
the total source flux in the 0.5--10 keV (from {\it Swift}/XRT data)
plus 15--50 keV (from {\it Swift}/BAT) energy bands to the source flux
in the 0.5--10 keV energy band to estimate the bolometric correction
factor throughout the decay phase of the outburst.

The observed dependence of this ratio on flux in the 0.5--10 keV band
is shown with filled blue points in Fig.~\ref{fig:bol}. To relate the
observed flux ratio and the actual bolometric correction factor
$K_{\rm bol}$ we used the broadband observations performed with the
{\it NuSTAR} observatory (marked by red crosses). The final bolometric
correction factor values are shown with the open circles in the same
figure, and are consistent with bolometric flux estimate by
\cite{2018A&A...613A..19D} using the {\it NuSTAR}, {\it Swift}/BAT and
MAXI data at the higher end of fluxes considered here. In order to get
a simple recipe for conversion of the observed {\it Swift}/XRT flux to
the bolometric one we approximated the observed dependence of the
correction factor on the observed flux in the 0.5--10 keV band with
the Gauss error function, shown with the grey line. In the following
analysis we apply this correction to all observational data and refer
to the bolometrically corrected fluxes and luminosities unless stated
otherwise. As can be seen from Fig.~\ref{fig:bol}, the bolometric
correction remains almost constant for low luminosities, so our
results are essentially unaffected by the uncertainties in its
reconstruction.

\section{Results}

\begin{figure}
\centering
\includegraphics[width=0.98\columnwidth, bb=25 150 575 710]{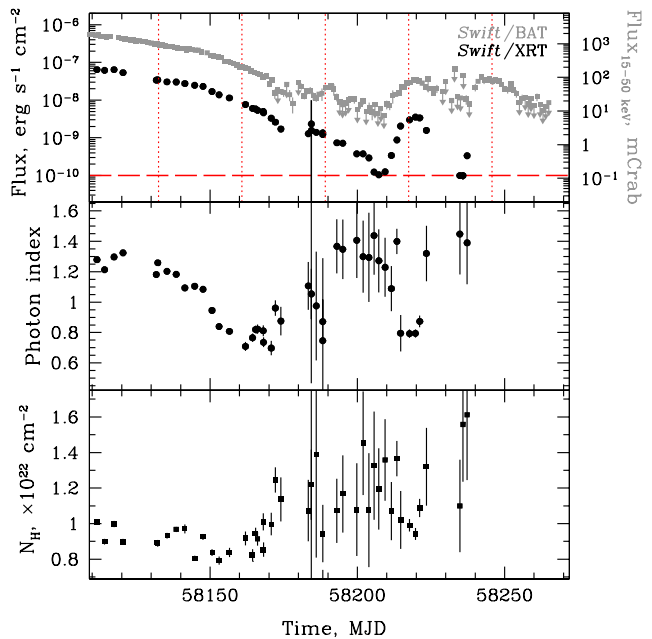}
\caption{{\it Top:} The bolometrically corrected light curve of \swift\ based on the
  {\it Swift}/XRT telescope data (black circles; left axis). Vertical
  red dotted lines represent moments of the periastron passage
  \citep{2018A&A...613A..19D}. Horizontal dashed line shows the upper
  limit on the threshold flux for the propeller effect onset. Grey
  squares show evolution of hard X-ray flux in the 15--50 energy range from {\it Swift}/BAT
  (right axis). Arrows correspond to the 2$\sigma$ upper limits.  {\it
    Middle:} The evolution of the spectral photon index with time.
  {\it Bottom:} The evolution of the hydrogen column density $N_{\rm
    H}$ with time.  }\label{fig:fluxlc}
\end{figure}

The bolometric and absorption corrected light curve of \swift\ during
the fading tail of the outburst is shown in the top panel of
Fig.~\ref{fig:fluxlc}.  A more or less gradual decrease of the flux is
observed until around MJD 58207 when it reached the minimal value of
$F_{\rm min}\simeq1.1\times10^{-10}$~\flux. After this point the
source started to rebrighten \citep{2018ATel11517....1R}, which was
followed again by a gradual decay. Another minimum with the same flux
level has been reached around MJD~58235, again followed by the
rebrightening.  This temporary flux increases did not allow us to
detect deeper minimum before the visibility was constrained by
Sun. Such rebrightenings during the fading phase of giant outbursts in
Be/X-ray pulsars systems were already reported by
\cite{2017A&A...605A..39T} for SMC~X-3.

 \begin{figure*}
 \centering
 \includegraphics[width=0.34\textwidth, bb=0 549 300 790, clip]{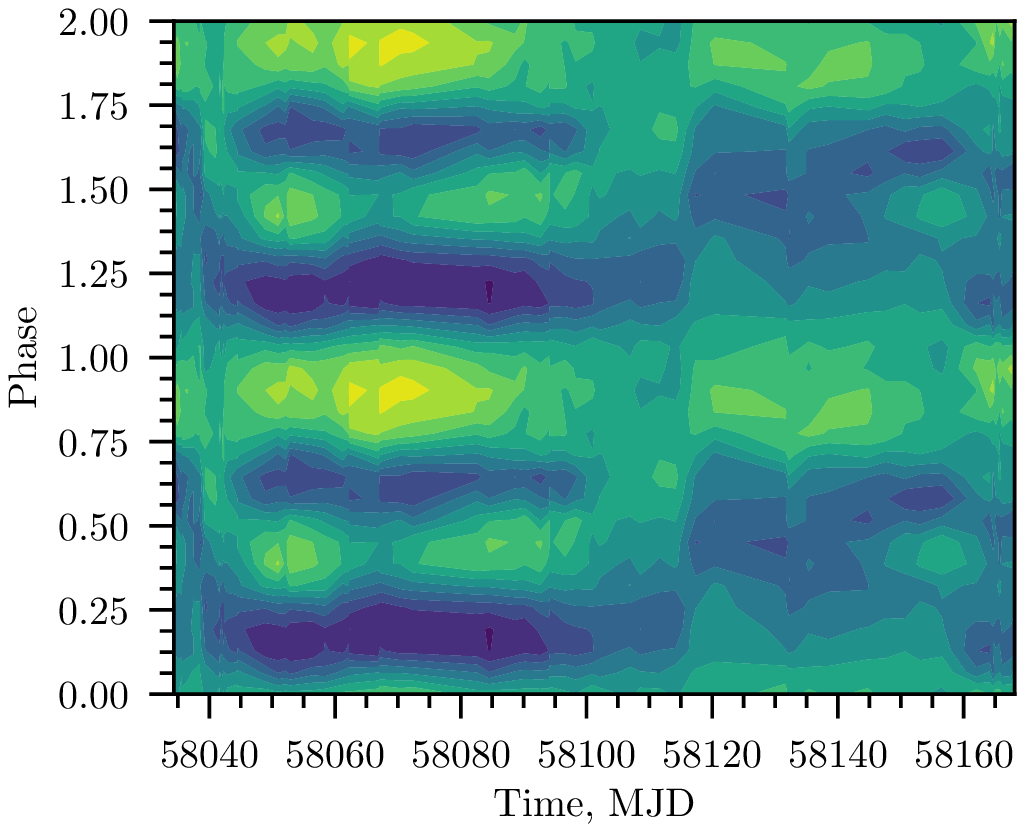}
 \hspace{0.1cm}
 \includegraphics[width=0.34\textwidth, bb=0 598 240 790, clip]{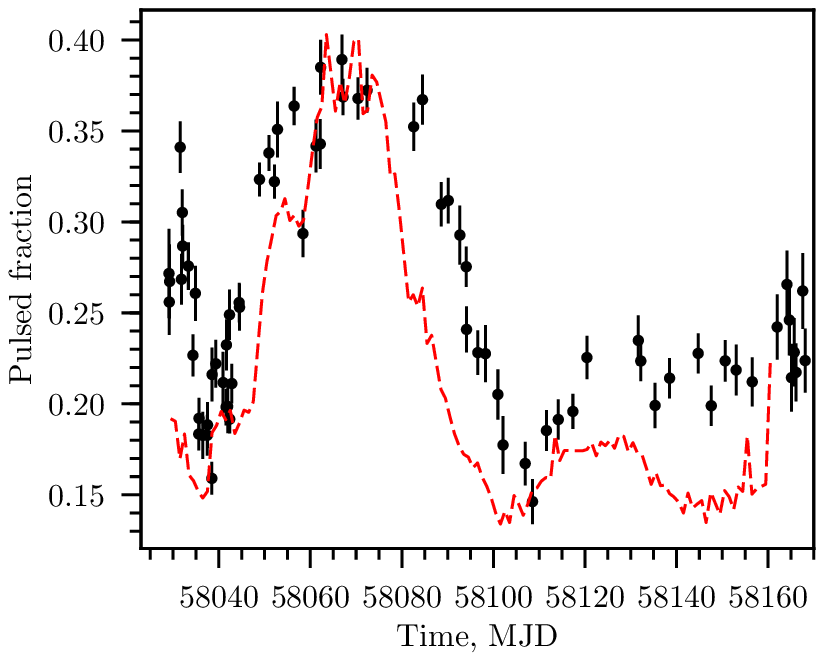}
 \hspace{0.1cm}
 \includegraphics[width=0.27\textwidth, bb=20 146 540 675]{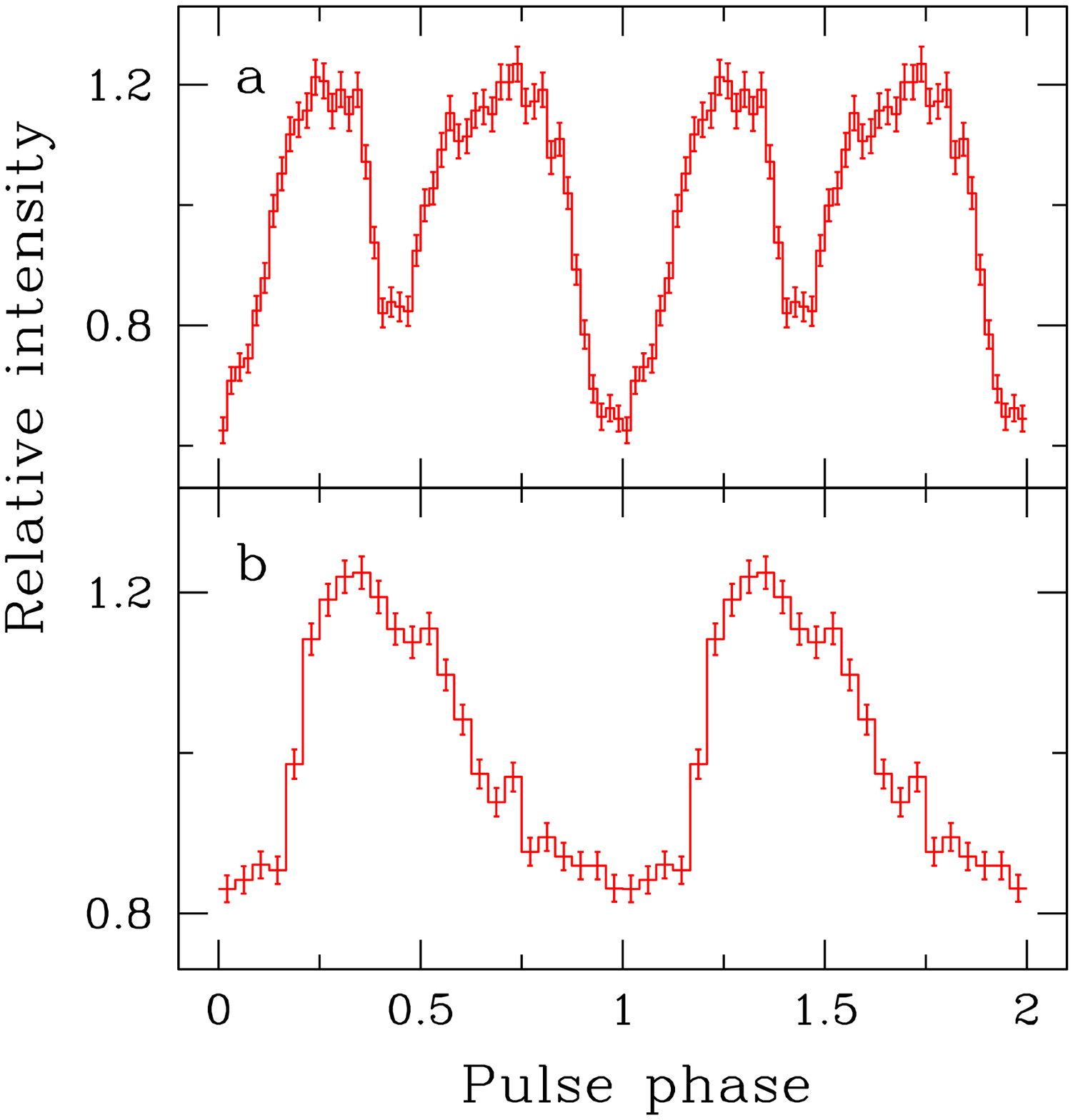}
 \caption{\emph{Left:} Evolution of the pulse profile of the source throughout the outburst
 as observed by \emph{Swift}/XRT in 0.8--10\,keV energy range. Slices along the ordinate give normalized intensity as function of pulse phase at a given time.
 Pulse profiles were roughly aligned to minimize the offset between pairs of consequent pulse profiles. \emph{Middle:} Observed
 pulsed fraction from the \emph{Swift}/XRT data as function of time (black points). Ratio of pulsed flux measured by \emph{Fermi}/GBM and total \emph{Swift}/BAT flux, 
 which gives indirect estimate of the pulsed fraction  in hard 15--50\,keV band is also shown (red line, arbitrarily scaled).
 \emph{Right:} Observed pulse profile shape in 0.8--10 keV band before (a) and after (b) transition around MJD~58110.}
 \label{fig:timing}
 \end{figure*}

The spectra of the source in the energy range covered by the XRT
telescope were fitted using the absorbed power law model ({\sc
  phabs*powerlaw} in the {\sc XSPEC} package). The best-fit parameters
demonstrate some moderate variability with the photon index being in
the range of 0.8--1.3. The evolution of the photon index with time is
shown in the middle panel of Fig.~\ref{fig:fluxlc}. Independently of
the flux level the spectra demonstrated moderate hydrogen column
density $N_{\rm H} \simeq (0.8-1.0)\times10^{22}$ cm$^{-2}$ (see
bottom panel of Fig.~\ref{fig:fluxlc}) comparable to the interstellar
absorption in the direction to the source, $N_{\rm
  H}\simeq0.9\times10^{22}$ cm$^{-2}$ \citep{2013MNRAS.431..394W}.

In order to probe changes in the geometrical and physical conditions
in the emitting regions in the vicinity of the NS, we studied also the
pulse profile evolution with the source intensity. As a result, a
drastic change of the pulse profile shape in the 0.5--10 keV energy
band was discovered around MJD~58100--58120 (see
Fig.\,\ref{fig:timing}).  In particular, the pulse profile at high
fluxes is double-peaked with nearly equal intensity of both peaks,
whereas at lower fluxes the profile has a single peak
structure. Similar changes occurred also during the rising part of the
outburst (around MJD~58040). Significant modifications of the pulse
profile shape is accompanied also by the variations of the pulsed
fraction.  The pulsed fraction\footnote{It was calculated as
  $\mathrm{PF}=(F_\mathrm{max}-F_\mathrm{min})/(F_\mathrm{max}+F_\mathrm{min})$,
  where $F_\mathrm{max}$ and $F_\mathrm{min}$ are the maximum and
  minimum flux in the pulse profile, respectively.} as a function of
time is shown in Fig.\,\ref{fig:timing}.  Note the strong changes on
MJD 58040--58055 and MJD 58090--58120. At higher energies the
observational coverage is limited, however, the source seems to
exhibit similar behaviour of the pulse profile and pulsed fraction also
in the hard band. Pulsed fraction variations along the outburst can be
traced in this case based on the comparison of the pulsed flux
measured by \emph{Fermi}/GBM in the 15--50 keV energy band and the
total {\it Swift}/BAT flux measured in the same energy band. As
illustrated in Fig.~\ref{fig:timing}, the pulsed flux fraction indeed
increases at high fluxes similarly to the pulsed fraction measured
directly by XRT. For the distance to the source of 7.3 kpc the
  transition luminosity can be estimated to be about $3\times10^{38}$ \lum.
It is important to emphasize that this change
is not associated with the transition to the propeller regime and
occurs at much higher luminosities.  However, as we discuss below, it
might also be relevant for the magnetic field strength estimation.

We were unable to obtain a robust phase coherent timing solution covering the
entire outburst due to rapid change of the short spin period and pulse profile
shape, comparatively large gaps between XRT pointings, and remaining
uncertainties in the orbital parameters. Therefore, we refrain from any
conclusions regarding the specific phase shifts between different observations,
however, strong changes of the observed pulse profile shape are apparent. To
illustrate the pulse profile evolution we show also two profiles well before
(MJD~58090) and after (MJD~58138) the transition in Fig.\,\ref{fig:timing}.
This transition is discussed in more detail in Section \ref{sec:discus}.

\section{Discussion}
\label{sec:discus}

Main goal of the work is to independently constrain the magnetic field
of the NS in \swift\ using the propeller effect. The propeller
luminosity $L_{\rm prop}$ is defined by the equality of the
magnetospheric radius ($R_{\rm m}$) to the co-rotation radius ($R_{\rm
  c}$).  Under the assumption of the Keplerian motion in the accretion
disc, matter can penetrate the magnetosphere and be accreted to the NS
only if $R_{\rm m}<R_{\rm c}$ \citep{1975A&A....39..185I}. In the
opposite case centrifugal barrier stops the accretion and abrupt drop
of the source luminosity should be observed.

The magnetospheric radius depends on the mass accretion rate and
magnetic field strength, so a simple equation linking the limiting
luminosity $L_{\rm prop}$ to the fundamental parameters of the NS can
be derived by equating the co-rotation and magnetospheric radii
\citep[see, e.g.][]{2002ApJ...580..389C}
\be\label{eq1} 
L_{\rm prop}(R) \simeq
\frac{GM\dot{M}_{\rm lim}}{R} \simeq 4 \times 10^{37} k^{7/2} B_{12}^2
P^{-7/3} M_{1.4}^{-2/3} R_6^5 \,\textrm{erg s$^{-1}$}, 
\ee 
where $R_6$ and $M_{1.4}$ are the NS radius and mass in units of $10^6$~cm and
1.4M$_\odot$, respectively, $B_{12}$ is the magnetic field strength at the
surface of the NS in units of $10^{12}$~G, $P$ is the pulsar rotational period in
seconds. Factor $k$ is required to take into account difference of the
magnetospheric radius in the case of disc accretion and classical Alfv\'en
radius ($R_{\rm m}=k \times R_{\rm A}$) and is usually assumed to be $k=0.5$
\citep{GL1978}. It was recently shown by \cite{2017A&A...608A..17T} that
transition to the propeller regime is possible only for pulsars with
relatively short pulse periods. From this point of view \swift, possessing
the period of $\sim9.7$ s, is a very good case study to make another step towards
the verification of theory of accretion from the ``cold disc''
\citep{2017A&A...608A..17T}.

As can be seen from Fig.~\ref{fig:fluxlc} no evidence of transition of
the source to the propeller regime was observed.  The lowest flux
level detected by the XRT telescope before \swift\ entered the
rebrightening phase is $F_{\rm min}\simeq1.1\times10^{-10}$~\flux. 

Using this value as an upper limit for the threshold flux $F_{\rm
  prop}$ it is possible to derive an upper limit for the magnetic
field strength.  For the distance of $d=7.3$~kpc and the coefficient
$k=0.5$, the propeller luminosity is $L_{\rm
  prop}<6.8\times10^{35}$~\lum\ and corresponding magnetic field
$B<6.2\times10^{12}$~G. This value is factor of two lower than the
value derived from the analysis of spin period derivative
\citep{2018A&A...613A..19D}. However, given the strong dependence of
the result on the assumed torque model and accretion disc effective
radius, we consider agreement satisfactory. In particular, it is
sufficient to assume $k\sim0.35$ \citep[see
  e.g.,][]{2017MNRAS.470.2799C} to increase the upper limit obtained
from non-detection of the transition to the propeller at $10^{13}$~G,
making it consistent with the accretion torque estimate. Note,
however, that improvement on the upper limit on the propeller
luminosity would deteriorate this agreement and thus imply presence of
an appreciable non-dipole component of the field.

Additional information about the magnetic field strength can be obtained from
the analysis of the pulse profile variations with luminosity. The sharp
variations of a beam pattern with accretion luminosity can be caused by the
 transition from the sub-critical regime of accretion to the
super-critical regime, when the radiation pressure becomes high enough to stop
matter above the NS surface \citep{1976MNRAS.175..395B,2012A&A...544A.123B,2015MNRAS.447.1847M}. In
this case, rather small variations of the mass accretion rate can result in
appreciable changes of the geometry of the emitting region and modification of
the observed pulse profile and pulsed fraction \citep{1973A&A....25..233G}.

The critical luminosity value $L_{\rm crit}$ depends on the magnetic
field strength at the NS surface \citep{2012A&A...544A.123B,2015MNRAS.447.1847M}. For a
``standard'' magnetic field $B\sim10^{12}$~G it is estimated at
$L_{\rm crit} \sim 10^{37}$~\lum, and increases for stronger magnetic
fields. If the observed variations of the pulse profile and pulsed
fraction in \swift\ are caused by the transition through the critical
luminosity, it points to the magnetic field of the order of $\gtrsim
10^{13}$~G at the surface.  This value is, again, marginally
compatible with the other estimates.  We emphasize, however, that the
transition to the propeller was actually not observed, so the
transition flux can be, in principle, significantly lower, which would
make the two estimates inconsistent with each other.  As discussed in
\cite{2017A&A...605A..39T}, this discrepancy can, however, still be
resolved assuming that the magnetic field of the NS has
non-negligible multipole component. It also indicates that local X-ray
radiation is dominated by extraordinary mode of polarization
characterized by a lower cross-section of interaction with the accreting
material \citep{2006RPPh...69.2631H,2016PhRvD..93j5003M}. 
We emphasize that to explore this possibility it is important to continue
monitoring of the source shall it enter another outburst.

\section{Conclusions}
\label{sec:conc}

Here we presented the results of the monitoring of the newly discovered
unique X-ray pulsar \swift\ with the {\it Swift}/XRT telescope in the tail of
its giant outburst. The source in fact is the brightest BeXRP in the Milky 
Way and belongs to the recently discovered family of ultraluminous X-ray pulsars. 
The main goal of the observational campaign was
to detect the transition of the pulsar to the ``propeller'' state and,
hence, to estimate the dipole component of the magnetic field of the
NS powering this source.  Unfortunately, the visibility constraints
did not permit us to observe the source down to the fluxes low enough
to detect such a transition. However, we were able to put a tight upper
limit on the propeller luminosity $L_{\rm prop}<6.8\times10^{35}$~\lum, and,
correspondingly, an upper limit on the dipole component of the NS
magnetic field strength $B\lesssim10^{13}$~G. This value is in
line with estimates obtained based on the observed spin-up rate of the
pulsar, and possible transition of the pulsar through the critical
luminosity at $L_{\rm crit}\sim3\times10^{38}$~\lum\ 
suggested by the observed
drastic change of the pulse profile shape around MJD~58110.
\swift\ can serve as a unique Galactic laboratory for studying
  physics of ultraluminous X-ray pulsars over the large dynamic range
  of luminosities.

\section*{Acknowledgements}
This work was supported by the Russian Science Foundation grant
14-12-01287 (SST, AAL, AAM). VD thank the Deutsches Zentrum for Luft-
und Raumfahrt (DLR) and Deutsche Forschungsgemeinschaft (DFG) for
financial support. We also express our thanks to the {\it Swift} ToO
team for prompt scheduling and executing of our observations.


\bibliographystyle{mnras}
\bibliography{allbib}

\bsp    
\label{lastpage}
\end{document}